# Impact of Zr substitution on the electronic structure of ferroelectric hafnia

Jinhai Huang,[1,#] Ge-Qi Mao,[1,#] Kan-Hao Xue,[1,2,*] Shengxin Yang,[1] Fan Ye,[1] Huajun Sun,[1,2] and Xiangshui Miao[1,2]

[1]School of Integrated Circuits, Huazhong University of Science and Technology, Wuhan 430074, China

[2]Hubei Yangtze Memory Laboratories, Wuhan 430205, China

*Corresponding Author, E-mail: xkh@hust.edu.cn (K.-H. Xue)

#These authors contributed equally.

## ABSTRACT

$HfO_2$-based dielectrics are promising for nanoscale ferroelectric applications, and the most favorable material within the family is Zr-substituted hafnia, *i.e.*, $Hf_{1-x}Zr_xO_2$ (HZO). The extent of Zr substitution can be great, and $x$ is commonly set to 0.5. However, the band gap of $ZrO_2$ is lower than $HfO_2$, thus it is uncertain how the Zr content should influence the electronic band structure of HZO. A reduced band gap is detrimental to the cycling endurance as charge injection and dielectric breakdown would become easier. Another issue is regarding the comparison on the band gaps between $HfO_2$/$ZrO_2$ superlattices and HZO solid-state solutions. In this work we systematically investigated the electronic structures of $HfO_2$, $ZrO_2$ and HZO using self-energy corrected density functional theory. In particular, the conduction band minimum of $Pca2_1$-$HfO_2$ is found to lie at an ordinary k-point on the Brillouin zone border, not related to any interlines between high-symmetry k-points. Moreover, the rule of HZO band gap variation with respect to $x$ has been extracted. The physical mechanisms for the exponential reduction regime and linear decay regime have been revealed. The band gaps of $HfO_2$/$ZrO_2$ ferroelectric superlattices are investigated in a systematic manner, and the reason why the superlattice could possess a band gap lower than that of $ZrO_2$ is revealed through comprehensive analysis.

# 1. Introduction

The ferroelectricity observed in doped hafnia (HfO$_2$) [1, 2] has great revived the research interests on nanoscale oxide ferroelectrics, because it can well maintain the spontaneous polarization ($P_S$) at film thicknesses below 10 nm. A meta-stable $Pca2_1$ polar orthorhombic phase (*o*-phase) has been acknowledged as the origin of such ferroelectricity. Compared with Al, Si, La, or Gd-doped hafnia, Hf$_{1-x}$Zr$_x$O$_2$ (HZO) appears to be even more promising since ferroelectricity is observed over a large range of *x* values [3]. Moreover, the ferroelectric crystallization temperature of HZO is lower than Si- or Al-doped hafnia [4–7], and can be controlled within the limit of state-of-the-art microelectronics flow-line. There are at least two key reasons why hafnia becomes the choice of nanoscale oxide-based ferroelectric material family. On the one hand, the coercive field ($E_C$) of ferroelectric hafnia or HZO is typically greater than 1 MV/cm [8–10], compared with Pb(Zr$_x$Ti$_{1-x}$)O$_3$, SrBi$_2$Ta$_2$O$_9$ and Bi$_{3.25}$La$_{0.75}$Ti$_3$O$_{12}$ whose $E_C$ values are below 100 kV/cm [11–14]. Hence, the data retention of hafnia-based ferroelectrics is excellent even for films with merely 5 nm – 10 nm thickness. On the other hand, the band gap of hafnia is around 6 eV [15], much larger than TiO$_2$-based ferroelectrics (3.0–3.5 eV) [16]. This enables a reasonable level of leakage current even in extremely thin films. Nevertheless, it is well-known that the band gap of ZrO$_2$ is lower than that of HfO$_2$, since the Zr-O bonds are not so strong as Hf-O bonds [17]. There is a natural issue as whether Zr substitution should deteriorate the leakage current of hafnia ferroelectrics, and to what extent thereafter.

In this respect, there are several questions that remain to be answered.
(i) Although monoclinic phase, tetragonal phase and amorphous phase of hafnia have been well studied, how large is the band gap of the *o*-phase compared with the others?
(ii) What exactly is the influence of Zr states in the HZO electronic structure?
(iii) What are the differences between the electronic structures of HfO$_2$/ZrO$_2$ superlattice and that of HZO solid-state solutions?

Since the *o*-phase is meta-stable and usually coexists with various other phases, experimental investigation of the electronic structures for the *o*-phase is difficult. Recently, Pavoni *et al.* [18]

studied Zr-substituted HfO$_2$ in the monoclinic phase (*m*-phase) and *o*-phase, using GGA+U, where GGA stands for generalized gradient approximation of density functional theory [19, 20] (DFT). The Hubbard *U* parameters [21] were fitted to experimental band gaps. Their GGA+U band diagram for Hf$_{0.5}$Zr$_{0.5}$O$_2$ shows the conduction band minimum (CBM) lying at the Γ point. There is a conceptual issue with the GGA+*U* approach applied to *o*-HfO$_2$, because the experimental band gap for single crystalline *o*-HfO$_2$ or *o*-Hf$_{0.5}$Zr$_{0.5}$O$_2$ is missing. The optimized *U* value has to fit other phases or mixed phases. In this work, we attempt to resolve these issues through *ab initio* calculations without fitting parameters. As electronic structures with accurate band gaps are required, conventional DFT within local density approximation [20, 22–24] (LDA) or GGA [25–30] is inadequate due to the well-known band gap underestimation problem [31]. Therefore, we have adopted the self-energy corrected GGA-1/2 method [32, 33] as well as hybrid functionals. And the paper is organized as follows. First, the computational method, involving the technique of band gap rectification, is introduced. Subsequently, electronic structures of bulk HfO$_2$, ZrO$_2$ and HZO in various phases are compared, emphasizing the trend of band gap variation with respect to the Zr content in HZO. Then, the *o*-phase HfO$_2$/ZrO$_2$ superlattice models are investigated, with variable layer thicknesses. It is demonstrated that the radius mismatch between Hf and Zr cations is relevant to the band gap variation in HZO solid-state solutions with respect to the Zr content, but the distinct ionicity of Hf/Zr cations could render very unusual electronic structures in HfO$_2$/ZrO$_2$ ferroelectric superlattices.

## 2. Computational

**2.1 Basic computational settings and structural models**

For the sake of accuracy, we have implemented the state-of-the-art projector augmented-wave [34, 35] (PAW) method, as implemented in the Vienna *Ab initio* Simulation Package [36, 37] (VASP). The independent electron approximation is achieved through DFT, and the exchange-correlation is treated either with GGA or hybrid functionals. For the former we use the Perdew-Burke-Ernzerhof (PBE) functional [26], while the screened exchange Heyd-Scuseria-Ernzerhof (HSE06) [38, 39] hybrid functional is adopted for the latter. The plane wave kinetic energy cutoff is set to 500 eV. The electrons considered as in the valence are: 2s and 2p for O; 4s, 4p, 4d and 5s for Zr; 5p, 5d and 6s for Hf. As the PAW potentials are of fully separable Kleinman-Bylander type [40], the 4s and 4p semi-core electrons are treated as valence electrons for Zr, in order to avoid the presence of ghost states [41]. Moreover, consideration of the semi-core states for Zr and Hf is important for accurately recovering the structural parameters in their compounds [42]. The force criterion during structural optimization is set to 0.01 eV/Å along each direction. The Brillouin zone sampling is achieved through equal-spacing Monkhorst-Pack [43] $k$-meshes. More detailed computational parameters are specified in **Supplementary Note 1**.

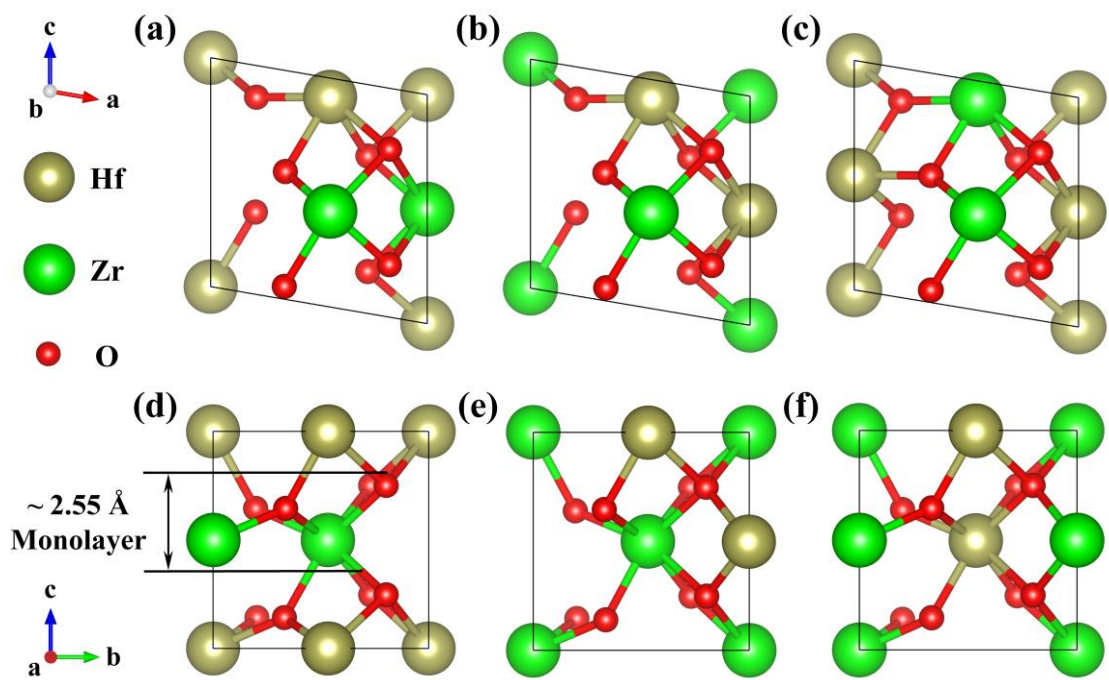

**Figure 1**. Atomic structures of various HZO models with $x$=0.5. (a) $m$-HZO in Model-I; (b) $m$-HZO in Model-II; (c) $m$-HZO in Model-III; (d) $o$-HZO in Model-I; (e) $o$-HZO in Model-II; (f) $o$-HZO in Model-III.

The $Pca2_1$-HfO$_2$ phase has two distinct O sites, $i.e.$ those with III-coordination (named O$_{III}$) and those with IV-coordination (name O$_{IV}$). It is the movement of O$_{III}$ anions that mainly accounts for the polarization reversal [44]. And there are multiples ways of Hf and Zr mixing in HZO. **Figure 1** illustrates the three fundamental mixing modes for both $m$-HZO and $o$-HZO. There relative energies are compared in **Table I**, where the overall most favorable phase ($m$-HZO Model-I) is set as the zero reference energy. It is obvious that the exact distribution of Hf/Zr has almost no impact on the energy of the monoclinic phase. Nevertheless, Model-I possesses the lowest energy among the orthorhombic $Pca2_1$ phases. In this model, Hf and Zr layers emerge alternately along the $c$-axis.

Experimentally, the $HfO_2$ and $ZrO_2$ layers are indeed grown through atomic layer deposition (ALD) alternately. Provided that a grain is crystallized in the *o*-phase with the polarization axis aligned vertically, then the resulting structure is highly similar to Model-I. Such conclusion is also consistent with a published work [45]. Hence, Model-I is adopted in this work for HZO.

**Table I**. Relative energies of various HZO models, in eV/f.u. unit (f.u. stands for an $Hf_{0.5}Zr_{0.5}O_2$ formular unit). The lowest energy model is set to zero energy.

|  | **Model-I** | **Model-II** | **Model-III** |
|---|---|---|---|
| ***m*-HZO** | 0 | $8.551 \times 10^{-5}$ | $3.232 \times 10^{-4}$ |
| ***o*-HZO** | 0.769 | 0.774 | 0.776 |

**2.2 Band gap rectification**

Accurately recovering the band gap of a semiconductor or an insulator is challenging for DFT [46]. The semi-local form of exchange in GGA is inaccurate and fails to cancel the spurious electron self-interaction [23, 47] completely. The consequence is manifested in several aspects. On the one hand, it violates the so-called Koopmans' compliant [48, 49] and produces electronic states that are too delocalized [50, 51]. On the other hand, the self-interaction error leads to over-small band gaps in semiconductors and insulators [23]. The HSE06 hybrid functional is a proper choice in solid-state calculations, since it introduces a part of Hartree-Fock non-local exchange, which is further screened to better reflect the solid-state situation. The improvement in exchange generally yields satisfactory band gaps. Nevertheless, the computational load of hybrid functionals is very heavy in solid-state calculations, and we in the mean time have to set up very large supercell for superlattice calculations. Hence, an efficient self-energy correction method, GGA-1/2 [32, 33, 52], is further attempted. For an insulator like $HfO_2$, it is inferred that the gap underestimation is mainly due to a too high level of valence band stemming from O states, which further stems from the unphysical electron self-interaction. Hence, through attaching an O "self-energy potential", which is attractive for electrons, the valence band can be lowered to the proper position, recovering the fundamental gap. Such self-energy potential is derived from atomic calculation from O, and should be properly trimmed before introducing into a solid (the overlapping of 1/*r* Coulomb tails leads to energy divergence) [32]. The cutoff radius is therefore a parameter, but it is obtained variationally to maximize the band gap, or to minimize the energy of the neutral ground state [32]. The method is explained in detail in several published works [32, 52–54]. There have been several computational works showing that GGA-1/2 works well for $HfO_2$ [55–57]. In this work, we shall systematically compare the electronics structures calculated using GGA-1/2 and HSE06, for $HfO_2$, $ZrO_2$ as well as HZO. The optimized cutoff radii in GGA-1/2 runs are given in **Supplementary Note 2**.

## 3. Results and Discussion

### 3.1 Electronic band structure of bulk *o*-phase

The electronic energy band diagrams of $HfO_2$ in monoclinic phase (*m*-phase, $P2_1/c$), tetragonal phase (*t*-phase, $P4_2/nmc$), *o*-phase and cubic phase (*c*-phase, $Fm\bar{3}m$) are compared in **Figures 2(a)-2(d)**. The continuous solid lines represent GGA-1/2 band structures, while the lowest conduction band as well as the topmost valence band derived through HSE06 are also marked in the figures using discrete circles. For reference, the corresponding Brillouin zones are illustrated in **Figure 3**. It follows that our GGA-1/2 band structures are in general consistent with HSE06 results, though the GGA-1/2 gaps are usually slightly larger than that of HSE06. The experimental band gap of $HfO_2$ is ~5.9 eV [15], in reasonable agreement with both GGA-1/2 and HSE06 results for *m*-$HfO_2$. In order not to miss the CBM and valence band maximum (VBM) points, we have also calculated the band gaps in terms of the Kohn-Sham eigenvalues (either GGA-1/2 corrected or not) from ultra-dense Monkhorst-Pack k-meshes. In case the as-derived band gap is smaller than that obtained from the band diagram (only considering high symmetry *k*-points and their inter-lines), it means the true CBM or VBM has been missed. For *m*-, *t*- and *c*-phases, the band diagrams shown in **Figure 2** already involve VBM and CBM, as expected. Nevertheless, the true GGA-1/2 band gap (~6.45 eV) of *o*-$HfO_2$ is smaller than any possible value extracted from the band diagram (the minimum apparent gap is 6.51 eV). The same phenomenon is observed in *o*-$Hf_{0.5}Zr_{0.5}O_2$. Hence, the electronic structure of *o*-$HfO_2$ is very special and deserves more in-depth investigation.

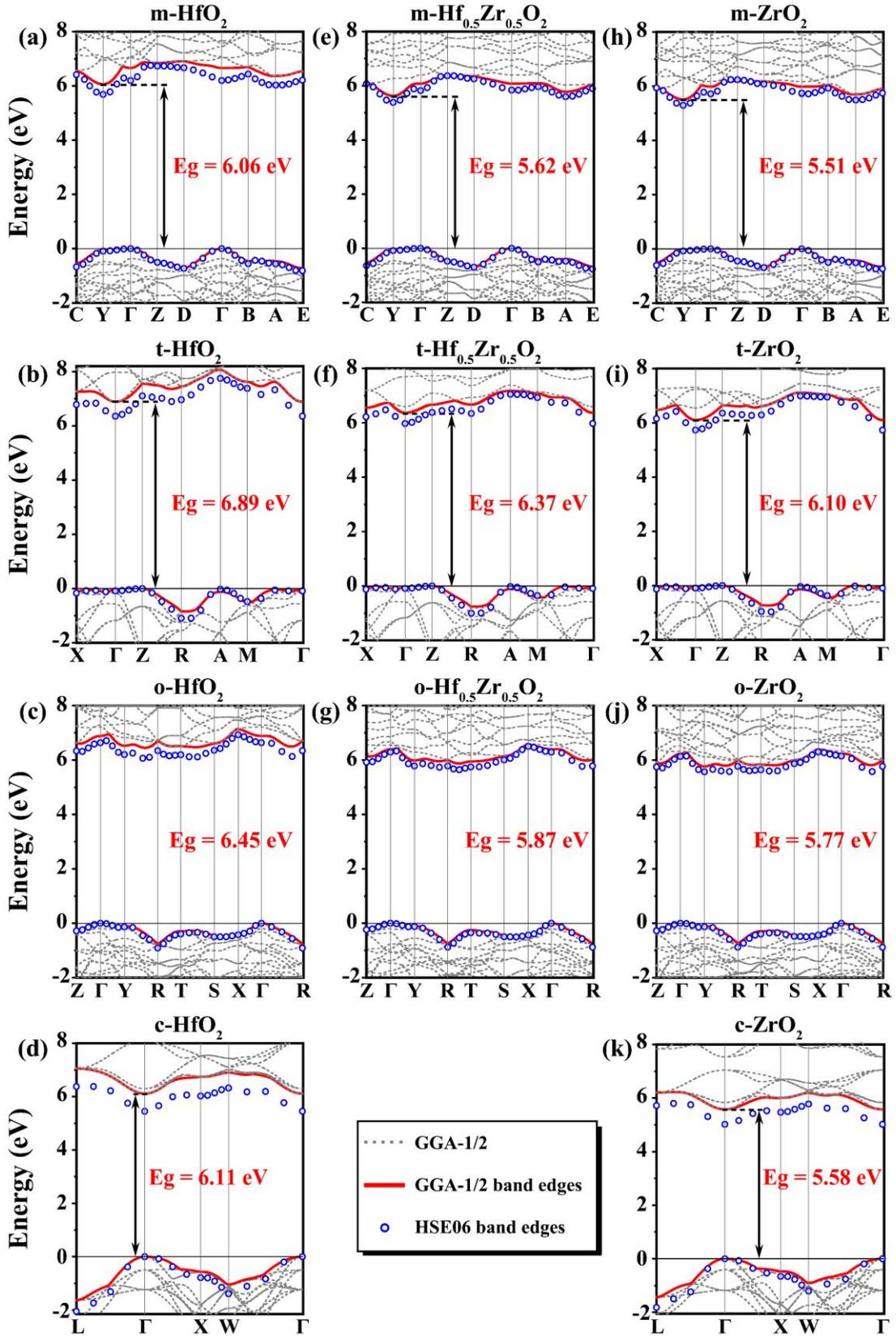

**Figure 2**. Electronic band diagrams calculated for HfO$_2$, HZO(Hf$_{0.5}$Zr$_{0.5}$O$_2$) and ZrO$_2$ for several phases (monoclinic *m*-, tetragonal *t*-, orthorhombic *o*-, and cubic *c*-), using GGA-1/2 and HSE06. (a) *m*-HfO$_2$; (b) *t*-HfO$_2$; (c) *o*-HfO$_2$; (d) *c*-HfO$_2$; (e) *m*-HZO; (f) *t*-HZO; (g) *o*-HZO; (h) *m*-ZrO$_2$; (i) *t*-ZrO$_2$; (j) *o*-ZrO$_2$; (k) *c*-ZrO$_2$.

To explore the true VBM and CBM locations, we have conducted guided search over the entire Brillouin zone of $o$-HfO$_2$, which aims at finding the highest eigenvalue of the occupied bands, as well as the lowest eigenvalue of the unoccupied bands. It turns out that the VBM is still at Γ, the Brillouin zone center. Nevertheless, the CBM lies on a plane Y-S-R-T on the Brillouin zone border. The exact CBM location has a coordination (0.275, 0.500, 0.348) in the reciprocal space, which lies somehow close to the line between Y and R, but shows obvious deviation. In **Figure 4**, the lowest unoccupied energy eigenvalues are plotted for the entire Y-S-R-T plane of the Brillouin zone, for both GGA and GGA-1/2 calculation results. It is clearly identified that both results predict the CBM at an ordinary location on the zone border, not subject to any special k-points or even their inter-lines. In the case of $o$-ZrO$_2$, the CBM is predicted at a similar ordinary k-point on the Y-S-R-T plane according to GGA calculations, similar to $o$-HfO$_2$. However, GGA-1/2 predicts the CBM at a high symmetry point Y. Actually, **Figures 4(c)-4(d)** clearly show two electron valleys on the Y-S-R-T plane. One is at Y and the other is at an ordinary location, which are competing with each other.

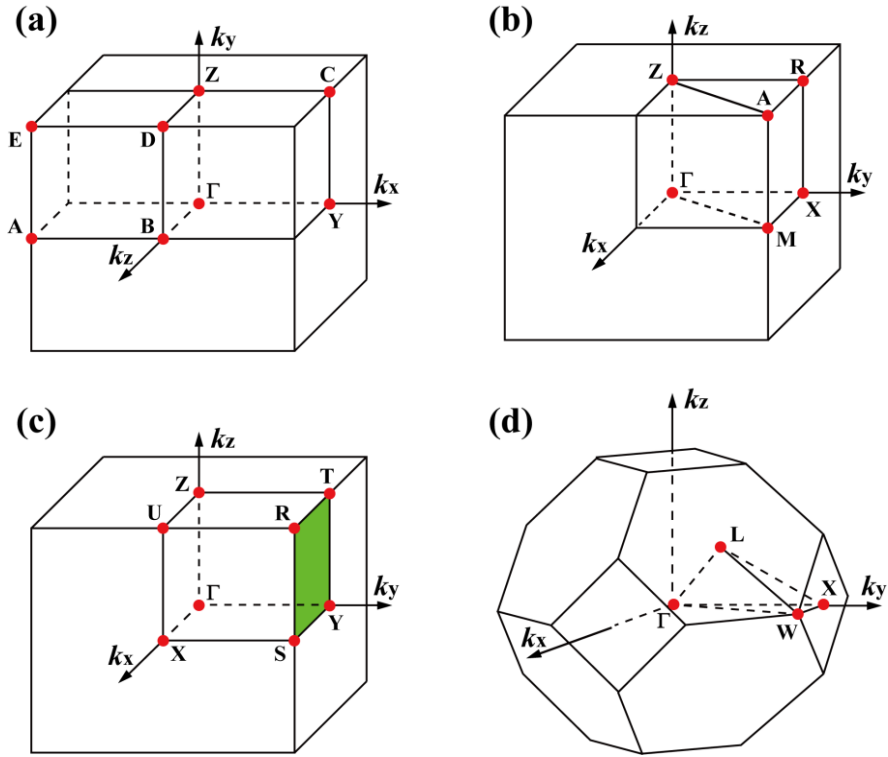

**Figure 3**. Brillouin zones and high-symmetry k points. (a) $m$-phase $P2_1/c$; (b) $t$-phase $P4_2/nmc$; (c) $o$-phase $Pca2_1$; (d) $c$-phase $Fm\bar{3}m$.

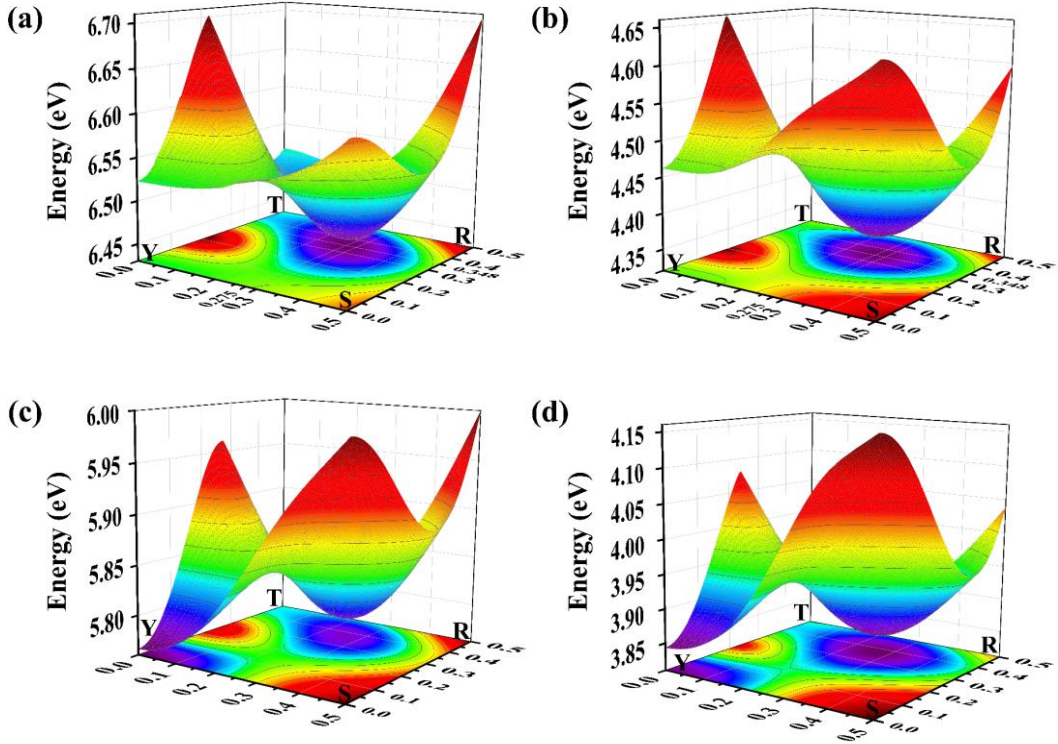

**Figure 4**. Detailed exploration on the conduction band edge in $o$-HfO$_2$ and $o$-ZrO$_2$ over a relevant border surface of the Brillouin zone. (a) $o$-HfO$_2$ calculated using GGA-1/2; (b) $o$-HfO$_2$ calculated using GGA; (c) $o$-ZrO$_2$ calculated using GGA-1/2; (d) $o$-ZrO$_2$ calculated using GGA.

**3.2 Band gap comparison among various phases**

The magnitudes of fundamental gaps follow the sequence of $t$-phase > $o$-phase > $c$-phase > $m$-phase. The GGA-1/2 predicted gap is 6.45 eV for $o$-phase HfO$_2$, which is noticeably larger than that of ordinary $m$-phase ($E_g$ = 6.06 eV). This implies that the intrinsic $o$-phase could resist leakage current better than the $m$-phase, though actually the leakage in hafnia mainly stems from grain boundaries [58]. The same trend of band gap ranking is observed in Hf$_{0.5}$Zr$_{0.5}$O$_2$ (**Figures 2(e)-2(g)**) and ZrO$_2$ (**Figures 2(h)-2(k)**) as well. For Hf$_{0.5}$Zr$_{0.5}$O$_2$, Model-I was adopted where Zr-layer and Hf-layer tend to appear alternately along the $c$-axis of the $Pca2_1$ phase. This is a potential benefit for HZO ferroelectric thin films in terms of experimental growth. Another encouraging fact lies in that, as just shown through calculations, the $o$-phase is the one with relatively large band gaps among the various common phases in the hafnia-zirconia system.

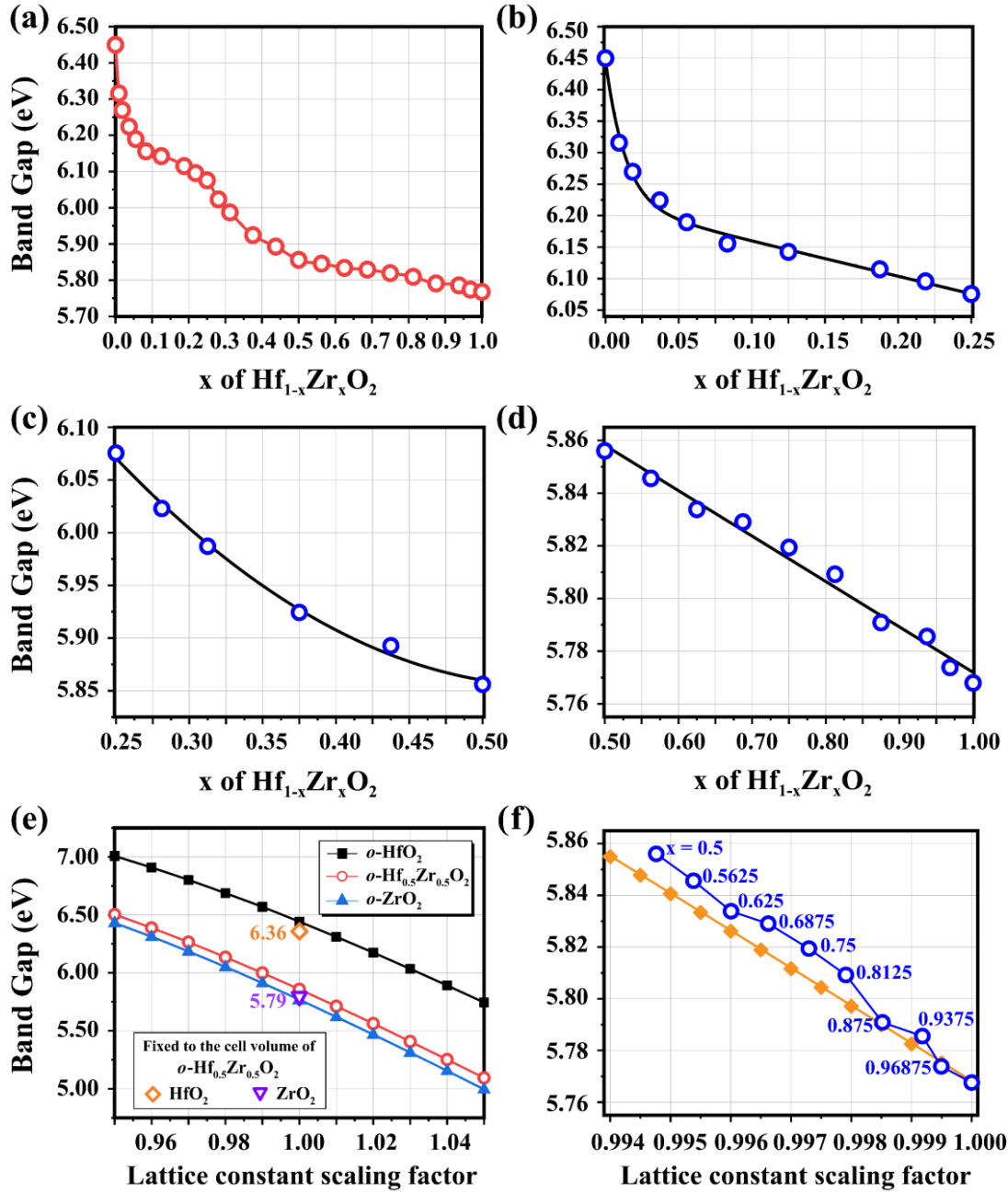

**Figure 5**. (a) Variation of GGA-1/2 band gap with respect to $x$ in $o$-Hf$_{1-x}$Zr$_x$O$_2$; (b) Curve fitting for the $x$ range of [0, 0.25]; (c) Curve fitting for the $x$ range of [0.25, 0.5]; (d) Curve fitting for the $x$ range of [0.5, 1]; (e) Band gap variation for $o$-HfO$_2$, $o$-ZrO$_2$ and $o$-Hf$_{0.5}$Zr$_{0.5}$O$_2$ under lattice scaling, with two additional band gaps of $o$-HfO$_2$ and $o$-ZrO$_2$ calculated at optimized $o$-Hf$_{0.5}$Zr$_{0.5}$O$_2$ unit cell volume marked; (f) Pure $o$-ZrO$_2$ band gaps at various lattice scaling factors, compared with that of equilibrium $o$-HZO with various $x$ values, whose effective scaling factors are marked (scaling factor is 1.000 when $x$ = 1).

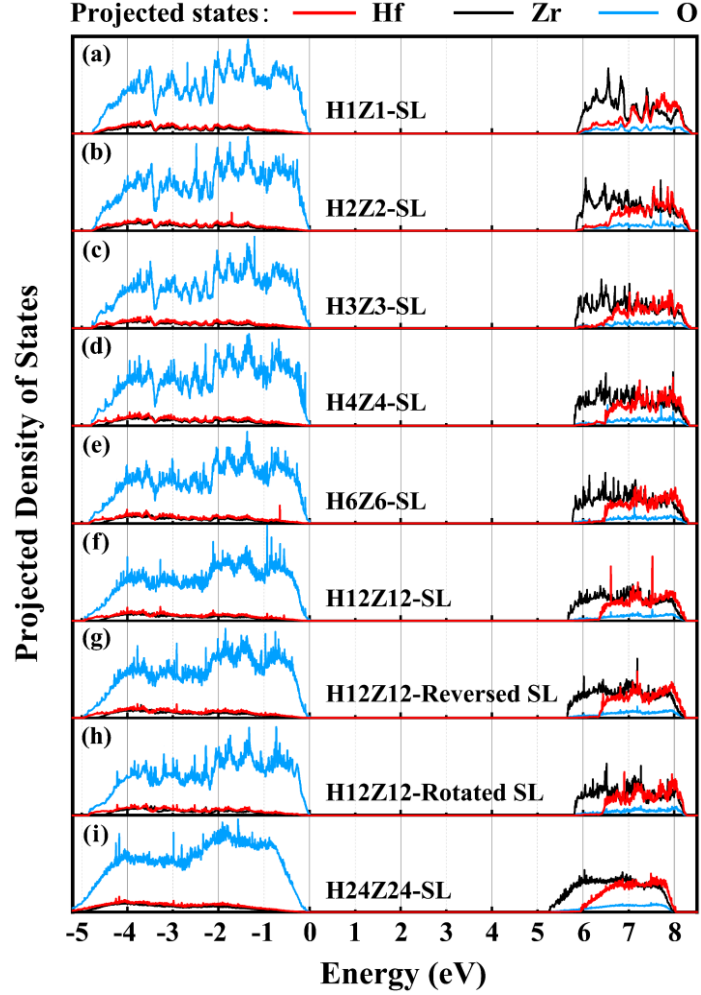

**Figure 6**. Partial density of states of various HZO SL models. Reversed superlattice stands for a model with polarization reversed with respect to the original model. The rotated superlattice has Hf/Zr/Hf/Zr… layers growing along the *b*-axis of the *Pca*2$_1$ structure, which is not a polar direction.

**3.3 Impact of Zr content on the band gap**

An interesting question is regarding the trend of gap variation upon substituting Zr for Hf. According to a comparison between *o*-HfO$_2$ (6.45 eV), *o*-Hf$_{0.5}$Zr$_{0.5}$O$_2$ (5.87 eV) and *o*-ZrO$_2$ (5.77 eV), it seems that the rule of gap decrease is not in a simple linear fashion. Therefore, we have set up a series of *o*-Hf$_{1-x}$Zr$_x$O$_2$ model supercells with *x* ranging from 0 to 1. **Figure 5(a)** demonstrates the variation of band gap ($E_g$) with respect to *x*, which clearly shows a sharp decrease when a tiny amount of Zr is substituted in HfO$_2$. The mathematical form of $E_g$ in HZO can be well fitted by an exponential function (supplemented with a linear function) for $x \leq 0.25$, in the unit of eV,

$$E_g(\text{HZO}) = -0.56x + 6.21577 + 0.30529 * exp\left(-\frac{x+0.00385}{0.01364}\right), \ 0 \leq x \leq 0.25$$

while **Figure 5(b)** confirms the quality of fitting. On the other hand, for *x* close to 1 it is observed that a linear fitting works well, as demonstrated in **Figure 5(d)**. The fitting result is ($E_g$(ZrO$_2$)=5.768 eV)

$$E_g(\text{HZO}) = E_g(\text{ZrO}_2) - 0.17243(x - 1) \ , \ 0.5 \leq x \leq 1$$

while the absolute value of the slope, 0.172, is small compared with the low-$x$ regime, but it is non-negligible. To explain these trends, we first carried out a partial density of states analysis for typical HZO samples with $x = 0.125$, $x = 0.5$ and $x = 0.875$. As illustrated in **Figure 6**, the CBM is dominated by Zr states in HZO. This is reasonable because the band gap of ZrO$_2$ is lower than that of HfO$_2$, thus a tiny doping of Zr could greatly reduce the band gap of HfO$_2$, consistent with the trend for $x \leq 0.25$. Nevertheless, this argument cannot explain the linear decrease of band gap for $0.5 \leq x \leq 1$. The sole physical picture of Zr dominating the conduction band edge may render the conclusion that the band gap will not change with $x$ as long as $x$ is sufficiently close to 1.

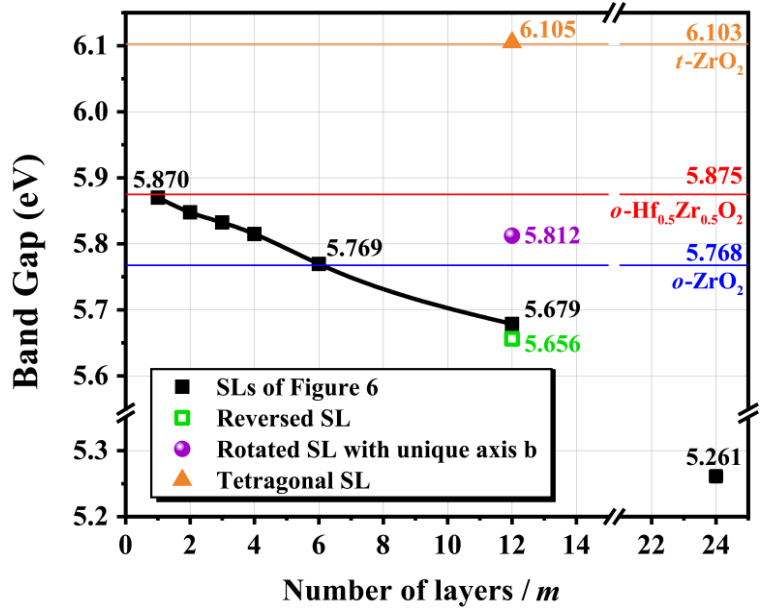

**Figure 7**. Band gap variation in HZO superlattices, with respect to the number of Hf/Zr layers in a period. Normally the Hf/Zr/Hf/Zr… layers pile up along the $c$-axis of the $Pca2_1$ structure. At 12-layer, several comparative models are introduced. Reversed superlattice stands for a model with polarization reversed with respect to the original model. The Rotated superlattice has Hf/Zr/Hf/Zr… layers growing along the $b$-axis of the $Pca2_1$ structure, which is not a polar direction. The tetragonal supercell means a HZO superlattice in the tetragonal paraelectric phase.

Hence, there are other reasons for the linear regime of $0.5 \leq x \leq 1$. We note that the lattice constants of HfO$_2$ and ZrO$_2$ are distinct, since Hf$^{4+}$ is a bit smaller than Zr$^{4+}$ (76 pm compared with 78 pm under VII coordination, according to Shannon ionic radii [59]), due to the lanthanide contraction [60]. Moreover, ZrO$_2$ contains more percentage of covalent bonding than HfO$_2$ [61], which tends to increase the size of the Zr cation. Zr substitution also plays a role in enlarging the lattice constants of HfO$_2$. Our test shows that the band gaps of both $o$-HfO$_2$ and $o$-ZrO$_2$ decrease when the lattice expands, as shown in **Figure 5(e)**. The HZO band gap varies in the same manner as its lattice (taking $x$=0.5 as an example) expands. Hence, more Zr substitution yields larger unit cells of HZO and lower band gaps. We have also conducted additional researches for better understanding of the volume effect. First of all, it is interesting to investigate the band gaps of $o$-HfO$_2$ and $o$-ZrO$_2$ when fixing to the cell volume of $o$-Hf$_{0.5}$Zr$_{0.5}$O$_2$. This means enlarging HfO$_2$ but compressing ZrO$_2$. The two additional data points in **Figure 5(e)** indeed show that the band gap of HfO$_2$ is reduced but that of ZrO$_2$ is enlarged, though the extent of change is very minor in either case. This is consistent with the fact that the volume effect is prominent only when $x$ is close to 1 (Zr-rich). Secondly, the linear variation of band gap observed for Zr-rich samples does not mean the volume effect must dominate. This is true only if the volume effect itself should account for the gap difference. To support of criticize this opinion, we have calculated the band gaps for a series of $o$-ZrO$_2$ models, each compared with that of HZO with a certain $x$ value, whose cell volume is the same as that of ZrO$_2$. **Figure 5(f)** demonstrates that the slopes of variation are nearly identical for the two cases, thus we confirm that the linear regime is indeed dominated by the cell volume argument.

The regime of $0.25 \leq x \leq 0.5$ is even more complicated as containing mechanisms from both low-$x$ regime and that of high-$x$ regime. The global effect is that the slope is steeper than the other two regimes. A quadratic fitting works,

$$E_\text{g}(\text{HZO}) = 6.5863 - 2.67335x + 2.44148x^2 \ , \ 0.25 \leq x \leq 0.5$$

While the band gap of HZO varies with $x$ in a complicated manner, it strictly falls within the range between that of pure $o$-ZrO$_2$ and pure $o$-HfO$_2$.

### 3.4 Comparison of HfO$_2$/ZrO$_2$ superlattice with solid-state solution

Considering the practical ALD growth, HZO may also conveniently be fabricated as a superlattice (SL). Provided that $m$ monolayers of HfO$_2$ and $m$ monolayers of ZrO$_2$ constitute a period along $c$-axis, the resulting SL will be named H$m$Z$m$-SL in this work. Among the various compositions,

H1S1-SL is actually the HZO solid-state solution, exactly the same as Model-I in **Figure 1(d)**. Hence, H1S1-SL is not really a SL, but this notation is still kept for better comparison with other true SLs. A series of HZO SLs with *m* ranging from 2 to 24 have been established. Their partial density of states analysis is given in **Figure 6**. It turns out that the band gap of HZO SL shrinks as *m* increases. The exact conduction band edges are dominated by Zr-states. To better compare the band gaps of these SLs with that of *o*-HfO$_2$ and *o*-ZrO$_2$, the variation of SL gap values with respect to *m* is further demonstrated in **Figure 7**. The most striking feature lies in that the SL band gap can be well below the value of *o*-ZrO$_2$ (5.768 eV), with *m*=12 and *m*=24 as typical examples. Such phenomenon is independent of the polarization direction, as another H12Z12-SL model with reversed polarization shows similar results (also marked in **Figure 7**). This discovery is rather surprising at first glance.

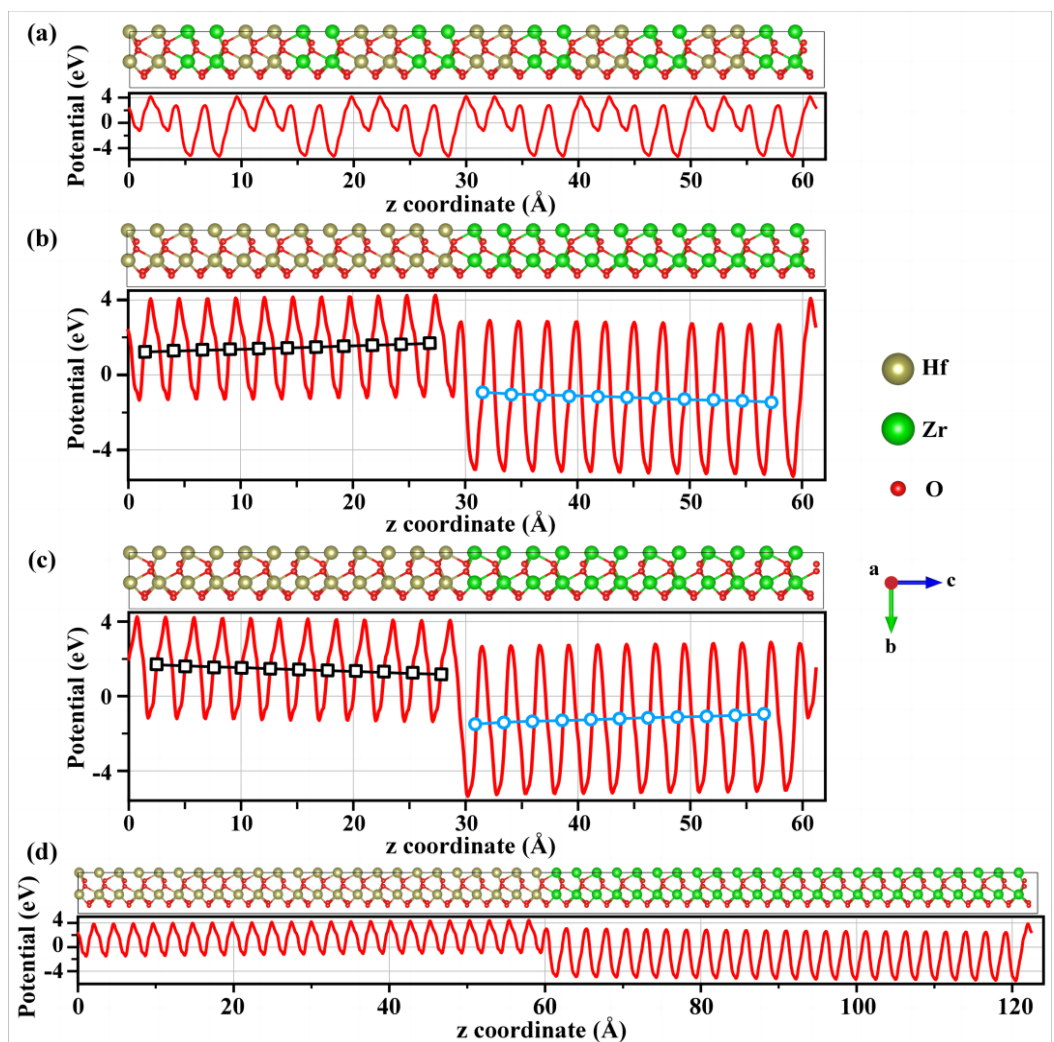

**Figure 8**. Distribution of local planar-averaged Hartree potential across HZO SL models. (a) H2Z2-SL; (b) H12Z12-SL; (c) Reversed H12Z12-SL; (d) H24Z24-SL.

To understand the band gap degradation at high *m*, we first attempted to measure the separation of Zr and Hf states ($\Delta E_c$) near the conduction band edge. As the electrons belong to the entire solid, there is no strict scheme to obtain atomic-decomposed states, and such measurement must be in a rough manner. Here, default Wigner-Seitz radii as specified in the PAW potentials have been used (Hf: 1.614 Å; Zr: 1.625 Å; O: 0.820 Å). And the separation is obtained from the Zr and Hf edges in the steep parts, which brings about some additional inaccuracy. However, the results are clear as both **Table II** and **Figure 6** show that the difference in $\Delta E_c$ between two adjacent models is merely in the range of 0.01 eV to 0.05 eV. This fails to account for the huge gap difference between, say, H6Z6-SL and H12Z12-SL (~0.1 eV difference in band gap). On the other hand, **Figure 6** clearly shows that the conduction band edges are less sharp in high-*m* SLs, especially for H24Z24-SL. Therefore, it is probable that some unknown mechanism spreads the states of the conduction band in energy.

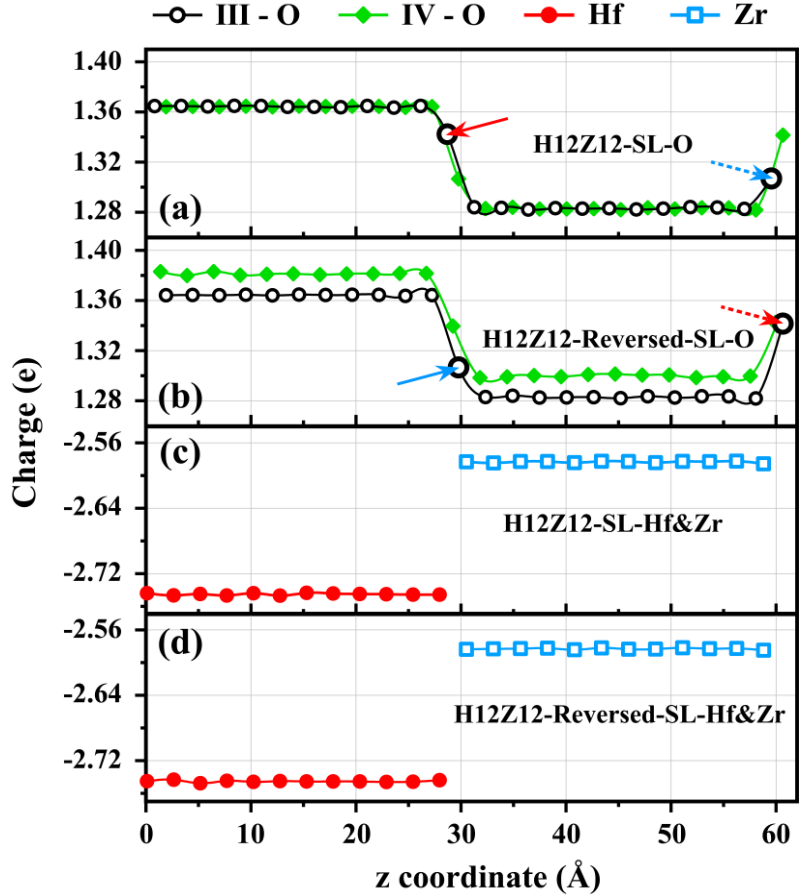

**Figure 9**. Bader charge analysis across the *c*-direction in an orthorhombic H12Z12-SL model.

Such mechanism is subsequently explored in terms electronic and chemical analyses. In **Figure 8**, the planar-averaged Hartree potentials along *c*-axis (with the unit of eV, in terms of electron potential energy) are demonstrated for several representative SL models. For H12Z12-SL, we explored both the original model and the polarization-reversed model. Provided that the average potential is calculated for each monolayer, it is possible to track the potential variation across the entire $HfO_2$ or $ZrO_2$ layer. As shown in **Figures 8(b)** and **8(c)**, their trends of potential variation are apparent opposite to each other. Nevertheless, one readily finds a common feature on the other hand, that the high potential energy point always lies at the $HfO_2/ZrO_2$ interface where the $O_{III}$ anions are closer to the Hf cations. Within each layer ($HfO_2$ or $ZrO_2$), a finite internal electric field is observed that spreads the conduction band states. Although some states go to higher energies while other go to lower energies, the band gap is counted from the lowest point of the conduction band (*i.e.* CBM), therefore the internal field is the reason for the band gap degradation in HZO SLs, especially for large *m* cases.

Table II. Rough estimation on the conduction band edge distance between Hf states and Zr states in *o*-HZO superlattices (SLs).

| Material | Distance (eV) | Material | Distance (eV) |
|---|---|---|---|
| H1Z1-SL | 0.278* | H12Z12-SL | 0.712 |
| H2Z2-SL | 0.700* | H12Z12-Reversed-SL | 0.715 |
| H3Z3-SL | 0.637 | H12Z12-Rotated-SL | 0.645 |
| H4Z4-SL | 0.691 | H24Z24-SL | 0.724 |
| H6Z6-SL | 0.715 | | |

*These values are not accurate because it is hard to identify the peak in the partial density of states.

One then naturally inquires why the potential energy should be higher when the interfacial $O_{III}$ anions are closer to the $HfO_2$ side. An analysis of the atomic charges is particularly useful towards this issue. Here we adopt the method by Bader [62], which was numerically improved by Tang *et al.* [63]. For H12Z12-SL models, either original or polarization reversed, the metal and oxygen Bader charges across the *c*-axis of the entire supercell are illustrated in **Figure 9**. The interfacial $O_{III}$ anions are specially marked using arrows. As the interfacial $O_{III}$ anion is closer to $HfO_2$ (solid arrow), it carries more negative charges than the $O_{III}$ anion closer to $ZrO_2$ on the other interface (dashed arrow). On the other hand, the Bader charges on Hf and Zr cations are relatively constant across each layer, and Hf is always more ionic than Zr, as confirmed by the results in **Figures 9(c) and 9(d)**. Hence, for each interface, the charge state is dominated by the amount of negative charges carried by O anions. Take the $HfO_2$ layer in H12Z12-SL model as an example. Its right interface has more negative charges than its left interface, and such difference can be regarded as a parallel-plate capacitor. Given that the surface charges are kept constant, then the voltage drop is proportional to the dielectric thickness. Hence, when the $HfO_2$ and $ZrO_2$ layers are thick as in H12Z12-SL, the internal potential variation spreads the conduction band states strongly. The polarity is such that the right interface ($O_{III}$ closer to Hf compared with Zr) possesses higher potential energy for electrons. The intrinsic reason is that $HfO_2$ is more ionic than $ZrO_2$ [61], thus $O_{III}$ obtains more electrons when it is adjacent to Hf, compared with Zr. Such argument of internal electric field polarity is consistent with both original and the reversed H12Z12 SL models.

## 4. Conclusion

The electronic structures of $HfO_2$, $ZrO_2$ and $Hf_{1-x}Zr_xO_2$ in their orthorhombic $Pca2_1$ phases and other common phases have been investigated using self-energy corrected density functional theory. For all materials, the magnitude of band gaps shows the same trend among different phases, *i.e.*, tetragonal > orthorhombic > cubic > monoclinic. The GGA-1/2 band gaps for $Pca2_1$-$HfO_2$ and $Pca2_1$-$ZrO_2$ are 6.45 eV and 5.77 eV, respectively. Surprisingly, the conduction band minimum of $Pca2_1$-$HfO_2$ lies neither on any high-symmetry k-point or the interlines between any two high-symmetry k-points in the Brillouin zone. It is located at an ordinary location on the Y-T-R-S surface of the zone border. Such behavior is very unusual among semiconductors and insulators, and may lead to over-estimated band gap calculation results, if the band diagram is plotted by simply connecting high symmetry k-points. The most favorable atomic structure of $Hf_{0.5}Zr_{0.5}O_2$ is the one with $HfO_2$ and $ZrO_2$ monolayers emerge alternately along the *c*-axis (polar direction) of the $Pca2_1$ phase. The band gap of $Pca2_1$-$Hf_{1-x}Zr_xO_2$ varies gradually from that of $Pca2_1$-$HfO_2$ to that of $Pca2_1$-$ZrO_2$ as *x* changes from 0 to 1, but the rule is quite complicated. For $x < 0.25$, the gap shows almost exponential decay, implying that a small amount of Zr substitution greatly reduces the band gap. For $0.5 < x < 1$, the gap varies almost linearly, mainly due to a volume expansion effect, because $ZrO_2$ is larger than $HfO_2$ in cell volume and these oxides show reduced band gaps when subject to expansion. On the other hand, when a ferroelectric superlattice is established from $HfO_2$ and $ZrO_2$ layers pile up alternately along the vertical polarization direction, the band gap decreases when the superlattice period extends, and can be even lower than that of *Pca*$2_1$-$ZrO_2$. This unexpected phenomenon is due to internal electric field stemming from the asymmetric interfaces of $ZrO_2$/$HfO_2$ and $HfO_2$/$ZrO_2$. When interfacial III-coordination O anions are closer to the Hf side, they carry more negative charges compared with those closer to the Zr side. The electric field spreads the Zr-states in the conduction band, shifting some states to lower energy levels, thus rendering the unexpectedly low band gap in $HfO_2$/$ZrO_2$ ferroelectric superlattices.

## Acknowledgement

This work was supported by the National Natural Science Foundation of China under Grant No. 61974049. The crystal structures were drawn using the VESTA software.

## Conflict of interest

The authors declare no competing interests.

## Supplementary information

"**Supplementary Note 1**. Detailed computational parameters" and "**Supplementary Note 2**. Cutoff radius scan for GGA-1/2" are added after the reference.

Supplementary Material for

# Impact of Zr substitution on the electronic structure of ferroelectric hafnia

Jinhai Huang,[1,#] Ge-Qi Mao,[1,#] Kan-Hao Xue,[1,2,*] Shengxin Yang,[1] Fan Ye,[1] Huajun Sun,[1,2] and Xiangshui Miao[1,2]

[1]School of Integrated Circuits, Huazhong University of Science and Technology, Wuhan 430074, China

[2]Hubei Yangtze Memory Laboratories, Wuhan 430205, China

*Corresponding Author, E-mail: xkh@hust.edu.cn (K.-H. Xue)

#These authors contributed equally.

## Supplementary Note 1. Detailed computational parameters

Parameters for the GGA calculations are as follows.

For Brillouin zone integration, we in general used the tetrahedron method with Blöchl correction. However, for band diagram calculations, the k points are in line mode, thus Gaussian smearing was used, with the width of the smearing fixed to 0.05 eV.

Sampling of the Brillouin zone was carried out through equal-spacing Monkhorst-Pack k-mesh. Detailed setting for various model supercells are listed in **Table S1**.

In calculating $Hf_{1-x}Zr_xO_2$ models with various Zr contents, there are supercells with 32 atoms or 108 atoms. The former was established by enlarging the *o*-$Hf_{0.5}Zr_{0.5}O_2$ unit cell to 2×2×2 in size, while the latter was established by enlarging the *o*-$Hf_{0.5}Zr_{0.5}O_2$ unit cell to 3×3×3 in size.

The stressed models were obtained by directing scaling the orthorhombic phases in their equilibrium structures. The k-mesh was also to 23×23×23.

Table S1. Supercell composition and k-mesh settings.

| Material | Number of atoms per cell | | | Size of k-mesh (relaxation) | Size of k-mesh (density of states) |
|---|---|---|---|---|---|
| | Hf | Zr | O | | |
| $m$-$HfO_2$ | 4 | 0 | 8 | | |
| $t$-$HfO_2$ | 2 | 0 | 4 | | |
| $o$-$HfO_2$ | 4 | 0 | 8 | | |
| $c$-$HfO_2$ | 4 | 0 | 8 | | |
| $m$-$Hf_{0.5}Zr_{0.5}O_2$ | 2 | 2 | 8 | | |
| $t$-$Hf_{0.5}Zr_{0.5}O_2$ | 1 | 1 | 4 | 12×12×12 | 23×23×23 |
| $o$-$Hf_{0.5}Zr_{0.5}O_2$ | 2 | 2 | 8 | | |
| $m$-$ZrO_2$ | 0 | 4 | 8 | | |
| $t$-$ZrO_2$ | 0 | 2 | 4 | | |
| $o$-$ZrO_2$ | 0 | 4 | 8 | | |
| $c$-$ZrO_2$ | 0 | 4 | 8 | | |
| $o$-$Hf_{0.99074}Zr_{0.00926}O_2$ | 107 | 1 | 216 | | |
| $o$-$Hf_{0.98148}Zr_{0.01852}O_2$ | 106 | 2 | 216 | | |
| $o$-$Hf_{0.96296}Zr_{0.03704}O_2$ | 104 | 4 | 216 | 2×2×2 | 3×3×3 |
| $o$-$Hf_{0.94444}Zr_{0.05556}O_2$ | 102 | 6 | 216 | | |
| $o$-$Hf_{0.91667}Zr_{0.08333}O_2$ | 99 | 9 | 216 | | |
| $o$-$Hf_{0.875}Zr_{0.125}O_2$ | 28 | 4 | 64 | | |
| $o$-$Hf_{0.8125}Zr_{0.1875}O_2$ | 26 | 6 | 64 | | |
| $o$-$Hf_{0.72125}Zr_{0.21875}O_2$ | 25 | 7 | 64 | | |
| $o$-$Hf_{0.75}Zr_{0.25}O_2$ | 24 | 8 | 64 | 6×6×6 | 9×9×9 |
| $o$-$Hf_{0.71875}Zr_{0.28125}O_2$ | 23 | 9 | 64 | | |
| $o$-$Hf_{0.6875}Zr_{0.3125}O_2$ | 22 | 10 | 64 | | |
| $o$-$Hf_{0.625}Zr_{0.375}O_2$ | 20 | 12 | 64 | | |
| $o$-$Hf_{0.5625}Zr_{0.4375}O_2$ | 18 | 14 | 64 | | |
| $o$-Hf$_{0.4375}$Zr$_{0.5625}$O$_2$ | 14 | 18 | 64 | | |
| $o$-Hf$_{0.375}$Zr$_{0.625}$O$_2$ | 12 | 20 | 64 | | |
| $o$-Hf$_{0.3125}$Zr$_{0.6875}$O$_2$ | 10 | 22 | 64 | | |
| $o$-Hf$_{0.25}$Zr$_{0.75}$O$_2$ | 8 | 24 | 64 | 6×6×6 | 9×9×9 |
| $o$-Hf$_{0.1875}$Zr$_{0.8125}$O$_2$ | 6 | 26 | 64 | | |
| $o$-Hf$_{0.125}$Zr$_{0.875}$O$_2$ | 4 | 28 | 64 | | |
| $o$-Hf$_{0.0625}$Zr$_{0.9375}$O$_2$ | 2 | 30 | 64 | | |
| $o$-Hf$_{0.03125}$Zr$_{0.96875}$O$_2$ | 1 | 31 | 64 | | |
| H1Z1-SL | | | | | |
| H2Z2-SL | | | | | |
| H3Z3-SL | | | | | |
| H4Z4-SL | 24 | 24 | 96 | 12×12×1 | 12×12×2 |
| H6Z6-SL | | | | | |
| H12Z12-SL | | | | | |
| H12Z12-Reversed-SL | | | | | |
| H12Z12-Rotated-SL | 24 | 24 | 96 | 12×1×12 | 12×2×12 |
| Tetragonal SL | 12 | 12 | 48 | 12×12×1 | 16×16×2 |
| H24Z24-SL | 48 | 48 | 192 | 12×12×1 | 9×9×1 |

Parameters for the HSE06 calculations are as follows.

The range separation parameter was set to 0.2 Å$^{-1}$, roughly corresponding to the default HSE06 setting. The amount of Hartree-Fock exchange mixing was 25% for the short-range part. A 4×4×4 equal-spacing k-mesh was used for structural relaxtion. The electronic band diagrams were obtained through a self-consistent HSE06 run, with k points coming from both the Monkhorst-Pack mesh (5×5×5) with finite weights, and from the k-lines that were set to zero weight.

## Supplementary Note 2. Cutoff radius scan for GGA-1/2

For $HfO_2$ and $ZrO_2$, the self-energy correction should be carried out on O anions only. However, the self-energy potential of O should be properly trimmed. The selected cutoff radius should maximize the band gap of the semiconductor/insulator. Figure S1 shows the variation of band gaps versus cutoff radius, for $HfO_2$, $ZrO_2$ and $Hf_{0.5}Zr_{0.5}O_2$ in their $Pca2_1$ phases. The power index $p$ of the cutoff function was chosen as the default value 8. For all three materials, the optimal cutoff radius was 2.7 bohr.

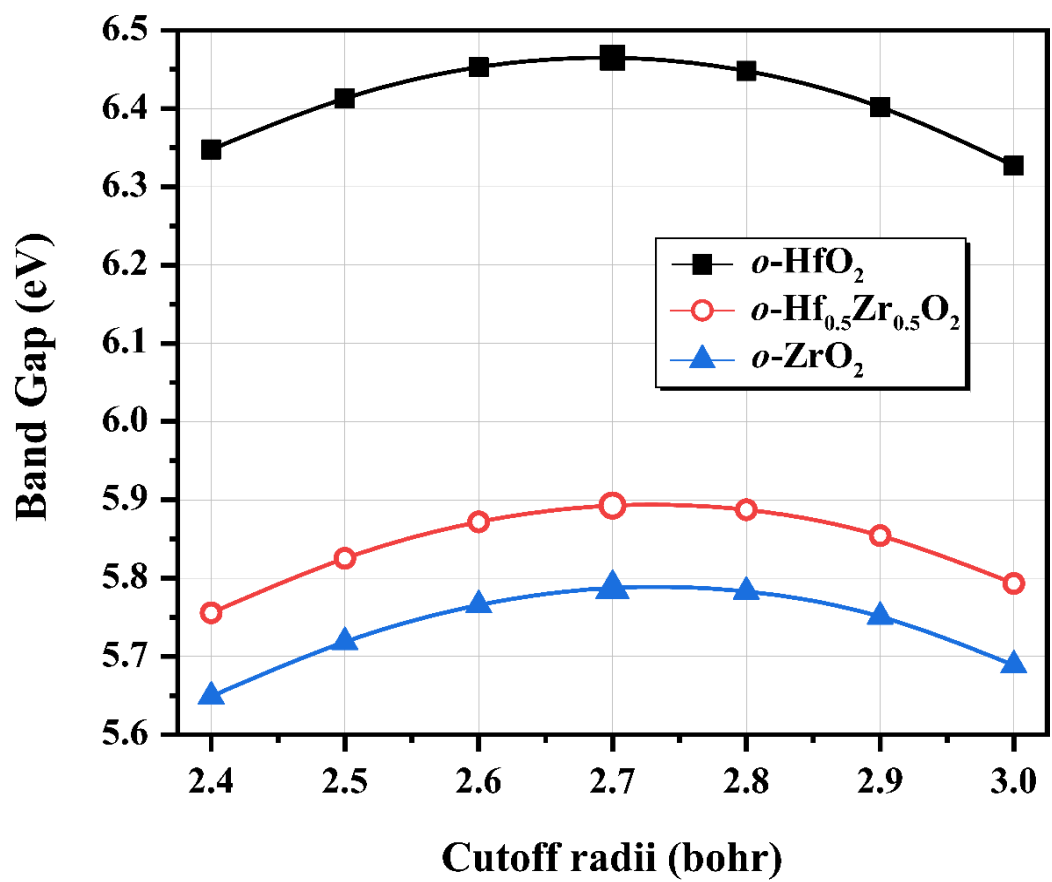

Figure S1. GGA-1/2 band gap versus self-energy potential cutoff radius.